\documentclass[sigconf, noacm]{acmart}
\AtBeginDocument{%
  }

\setcopyright{none} 
\renewcommand\footnotetextcopyrightpermission[1]{}

\usepackage{algorithmic} 
\usepackage{graphicx} 
\usepackage{textcomp}
\usepackage{multirow} 
\usepackage{makecell} 
\usepackage{url} 
\usepackage{enumitem} 

\newcommand{\cmark}{\checkmark}
\newcommand{\xmark}{\(\times\)}

\usepackage{listings}
\usepackage{calc}

\newsavebox{\loggroupbox}   
\newsavebox{\logwhitebox}   

\definecolor{logframe}{HTML}{8FBCE6}   
\definecolor{logbg}{HTML}{EBF6FB}      
\definecolor{myblue}{HTML}{4A78A8}  
\definecolor{myred}{HTML}{B35A5A}   
\definecolor{logwhite}{HTML}{FFFFFF}   

\lstdefinestyle{logwhite}{
  basicstyle=\ttfamily\footnotesize,
  breaklines=true,
  columns=fullflexible,
  keepspaces=true,
  showstringspaces=false,
  frame=none,
  aboveskip=0pt,
  belowskip=0pt,
  xleftmargin=0pt,
  xrightmargin=0pt,
  backgroundcolor=\color{logwhite},
  upquote=true
}

\definecolor{promptframe}{HTML}{8FBCE6}  
\definecolor{promptbg}{HTML}{EBF6FB}     
\definecolor{prompttitlebg}{HTML}{D6E7F6}

\newcommand{\promptbox}[2]{%
  \par\smallskip\noindent
  \begingroup
  \setlength{\fboxsep}{0pt}%
  \fcolorbox{promptframe}{promptbg}{%
    \begin{minipage}{\linewidth}

      \noindent\colorbox{prompttitlebg}{%
        \parbox{\linewidth}{%
            \vspace{4pt}%
            \hspace*{6pt}\textbf{#1}\hspace*{6pt}%
            \vspace{4pt}%
          }%
        }
      \par
      \begingroup
      \setlength{\parindent}{0pt}%
      \hspace*{6pt}%
      \begin{minipage}{\dimexpr\linewidth-12pt\relax}
        \vspace{4pt}
        {\normalsize\ttfamily #2}\par
        \vspace{4pt}%
      \end{minipage}%
      \endgroup

    \end{minipage}%
  }%
  \endgroup
  \par\smallskip
}

\newsavebox{\rqboxbox}

\newlength{\rqboxpadlr}   
\newlength{\rqboxpadtb}   
\newlength{\rqboxrule}    

\newenvironment{rqbox}{%
  \par\addvspace{6pt}
  \noindent
  \setlength{\fboxrule}{\rqboxrule}%
  \setlength{\fboxsep}{\rqboxpadlr}
  \begin{lrbox}{\rqboxbox}%
    \begin{minipage}{\dimexpr\linewidth-2\fboxsep-2\fboxrule\relax}%
      \normalfont\normalsize
      \setlength{\parskip}{0pt}%
      \setlength{\parindent}{0pt}%
      \kern\rqboxpadtb 
      \ignorespaces
}{%
      \unskip\kern\rqboxpadtb 
    \end{minipage}%
  \end{lrbox}%
  \fcolorbox{black!25}{gray!5}{\usebox{\rqboxbox}}%
  \par
}

\begin{document}

\title{LogSieve: Task-Aware CI Log Reduction for Sustainable LLM-Based Analysis}

\author{Marcus Emmanuel Barnes}
\orcid{0009-0000-7571-9085}
\affiliation{%
  \department{Faculty of Information}  
  \institution{University of Toronto}
  \city{Toronto}
  \state{Ontario}
  \country{Canada}
}
\email{marcus.barnes@utoronto.ca}

\author{Taher A. Ghaleb}
\orcid{0000-0001-9336-7298}
\affiliation{%
  \department{Department of Computer Science} 
  \institution{Trent University}
  \city{Peterborough}
  \state{Ontario}
  \country{Canada}}
\email{taherghaleb@trentu.ca}

\author{Safwat Hassan}
\orcid{0000-0001-7090-0475}
\affiliation{%
  \department{Faculty of Information}
  \institution{University of Toronto}
  \city{Toronto}
  \state{Ontario}
  \country{Ontario}
}
\email{safwat.hassan@utoronto.ca}

\renewcommand{\shortauthors}{Barnes et al.}

\begin{abstract}
Logs are essential for understanding Continuous Integration (CI) behavior, particularly for diagnosing build failures and performance regressions. Yet their growing volume and verbosity make both manual inspection and automated analysis increasingly costly, time-consuming, and environmentally costly. While prior work has explored log compression, anomaly detection, and LLM-based log analysis, most efforts target structured system logs rather than the unstructured, noisy, and verbose logs typical of CI workflows.

We present \textit{LogSieve}, a lightweight, \emph{RCA-aware and semantics-preserving} log reduction technique that filters low-information lines while retaining content relevant to downstream reasoning. Evaluated on CI logs from 20 open-source Android projects using GitHub Actions, \textit{LogSieve} achieves an average 42\% reduction in lines and 40\% reduction in tokens with minimal semantic loss.  This pre-inference reduction lowers computational cost and can proportionally reduce energy use (and associated emissions) by decreasing the volume of data processed during LLM inference.

Compared with structure-first baselines (\textit{LogZip} and random-line removal), \textit{LogSieve} preserves much higher semantic and categorical fidelity (Cosine = 0.93, GPTScore = 0.93, 80\% exact-match accuracy).  Embedding-based classifiers automate relevance detection with near-human accuracy (97\%), enabling scalable and sustainable integration of semantics-aware filtering into CI workflows.  \textit{LogSieve} thus bridges log management and LLM reasoning, offering a practical path toward greener and more interpretable CI automation.
\end{abstract}

\ccsdesc[500]{Software and its engineering~Software testing and debugging}
\ccsdesc[300]{Software and its engineering~Software maintenance tools}
\ccsdesc[300]{Computing methodologies~Information extraction}

\keywords{Continuous Integration, Log reduction, Large Language Models, LLMs, Semantic fidelity, Sustainability, Android}

\makeatletter
\renewcommand{\shortauthors}{}%
\renewcommand{\shorttitle}{}%
\let\@acmConference\@empty
\let\@acmYear\@empty
\let\@acmISBN\@empty
\let\@acmDOI\@empty
\let\@acmPrice\@empty

\fancypagestyle{emptyheadstyle}{%
  \fancyhf{}
  \fancyhead[L]{}
  \fancyhead[C]{}
  \fancyhead[R]{}
  \fancyfoot[C]{\thepage}
  \renewcommand{\headrulewidth}{0pt}%
  \renewcommand{\footrulewidth}{0pt}%
}
\pagestyle{emptyheadstyle}
\thispagestyle{emptyheadstyle}

\def\@copyrightspace{}%
\makeatother

\maketitle

\section{Introduction}
\label{sec:Introduction}

Logs are indispensable to Continuous Integration (CI) workflows, supporting tasks such as failure diagnosis, pipeline debugging, and observability. Prior work has shown that CI outcomes (e.g., breakages) interact with pipeline characteristics and their evolution, reinforcing the need for scalable diagnostics~\cite{ghaleb2022studying}. Yet their growing volume and verbosity make both manual inspection and automated analysis costly, time-consuming, and increasingly energy-intensive at scale~\cite{wu2022sustainable,he2021survey,gholamian2021comprehensive}. These challenges are particularly acute in ecosystems like GitHub Actions, where logs are voluminous, noisy, and loosely structured. The problem is even more pronounced in Android projects, whose multi-stage build and testing pipelines produce heterogeneous, verbose logs dominated by low-information content~\cite{ghaleb2025ci}. Prior studies have shown that Android apps often contain less-maintained logs and are more complex to parse~\cite{harty2021logging,zhu2019tools}, suggesting these problems carry over to their CI workflows.

The increasing adoption of large language models (LLMs) in software engineering has enabled a wide range of automation and reasoning tasks, but has also raised concerns about inference cost, latency, and environmental impact due to their sensitivity to input length~\cite{hou2024large}. In the context of CI, recent studies have explored the use of LLMs to automate tasks such as generating CI configurations~\cite{chomkatek2025decoding,ghaleb2025llm4ci} and migrating pipelines across CI services~\cite{hossain2025cigrate}. LLMs are also increasingly used for failure and anomaly detection with explanation~\cite{wang2024logexpert}, where CI logs serve as the primary input. In these settings, inference cost and latency scale with log length, meaning that every redundant line increases computational, financial, and carbon overhead.
Existing log-reduction methods focus primarily on \emph{structure-first} goals—compression, deduplication, or template clustering—to save storage or enable indexing~\cite{zhu2019logzip,li2024logshrink,he2017drain,du2017spell}. Such techniques improve infrastructure efficiency but overlook the \emph{task relevance} of individual lines for downstream reasoning. In contrast, a \emph{task-aware} perspective prioritizes semantic fidelity: retaining information essential for diagnosing and categorizing failures while discarding contextual noise, consistent with task-aware reduction strategies explored in other LLM pipelines~\cite{barnes2025task}. This shift from structure-first to semantics-first reduction is crucial for sustainable, interpretable CI automation, where reducing token budgets directly lowers both inference cost and environmental impact~\cite{Koenig2025,Naumann2011,schwartz2020green}.
In this paper, we focus on root-cause analysis (RCA) and failure categorization for GitHub Actions CI logs. Accordingly, task relevance is defined in terms of preserving the information necessary to support these diagnostic activities.

We present \textit{LogSieve}, a lightweight, RCA-aware log reduction technique that filters out low-information lines while retaining content relevant to downstream automation. Unlike compression-based approaches such as LogZip, \textit{LogSiev}e performs \textit{pre-inference, semantics-aware} reduction to minimize token count without compromising interpretability. The goal is to preserve diagnostic fidelity while reducing redundant computation, enabling scalable and sustainable CI analytics.

\begin{figure}[htbp]
  \centering
  \begin{lrbox}{\loggroupbox}
    \begin{minipage}{0.98\linewidth}
    \setlength{\fboxsep}{4pt}
      \noindent\colorbox{myblue}{\parbox{\dimexpr\linewidth-8pt\relax}{\vspace{1pt}\color{white}\footnotesize\textbf{Type: The executed command}\vspace{1pt}}}\par
      \begin{lrbox}{\logwhitebox}
        \begin{minipage}{\dimexpr\linewidth-8pt\relax}
\begin{lstlisting}[style=logwhite]
2025-05-01T04:19:28.9669135Z [command]/usr/local/lib/android/sdk/cmdline-tools/16.0/bin/sdkmanager tools
\end{lstlisting}
        \end{minipage}
      \end{lrbox}
      \noindent\colorbox{logwhite}{\usebox{\logwhitebox}}\par\vspace{2pt}

      \noindent\colorbox{myblue}{\parbox{\dimexpr\linewidth-8pt\relax}{\vspace{1pt}\color{white}\footnotesize\textbf{Type: Systematic process-related tasks}\vspace{1pt}}}\par
      \begin{lrbox}{\logwhitebox}
        \begin{minipage}{\dimexpr\linewidth-8pt\relax}
\begin{lstlisting}[style=logwhite]
2025-05-01T04:19:29.6442057Z Loading package information...
\end{lstlisting}
        \end{minipage}
      \end{lrbox}
      \noindent\colorbox{logwhite}{\usebox{\logwhitebox}}\par\vspace{2pt}

      \noindent\colorbox{myred}{\parbox{\dimexpr\linewidth-8pt\relax}{\vspace{1pt}\color{white}\footnotesize\textbf{Type: Task Progress}\vspace{1pt}}}\par
      \begin{lrbox}{\logwhitebox}
        \begin{minipage}{\dimexpr\linewidth-8pt\relax}
\begin{lstlisting}[style=logwhite]
2025-05-01T04:19:29.7185866Z [ ] 3% Loading local repository...
\end{lstlisting}
        \end{minipage}
      \end{lrbox}
      \noindent\colorbox{logwhite}{\usebox{\logwhitebox}}\par\vspace{2pt}

      \noindent\colorbox{myblue}{\parbox{\dimexpr\linewidth-8pt\relax}{\vspace{1pt}\color{white}\footnotesize\textbf{Type: Error message}\vspace{1pt}}}\par
      \begin{lrbox}{\logwhitebox}
        \begin{minipage}{\dimexpr\linewidth-8pt\relax}
\begin{lstlisting}[style=logwhite]
2025-05-01T04:23:19.7676299Z Execution failed for task ':app:compileOpenReleaseUnitTestJavaWithJavac'.
\end{lstlisting}
        \end{minipage}
      \end{lrbox}
      \noindent\colorbox{logwhite}{\usebox{\logwhitebox}}\par\vspace{0pt}

    \end{minipage}
  \end{lrbox}

  \setlength{\fboxsep}{4pt} 
  \fcolorbox{logframe}{logbg}{\usebox{\loggroupbox}}
  \Description{Example Android CI log lines with log types indicated and whether they are relevant or irrelevant for root-cause analysis.}
  \caption{Examples of LogSieve reductions on Android CI log lines. RCA relevance: red = irrelevant, blue = relevant.}
  \label{fig:logsieve-examples}
\end{figure}

To summarize, this paper makes the following contributions:
\begin{itemize}
    \item We propose \textit{LogSieve}, a \emph{RCA-aware, semantics-preserving} log reduction technique that filters low-information lines while retaining content relevant to failure diagnosis, bridging the gap between structure-first compression and LLM reasoning.
    \item We evaluate \textit{LogSieve} on CI logs from 20 open-source Android projects using GitHub Actions, achieving an average 42\% reduction in lines and 40\% reduction in tokens with minimal loss of diagnostic fidelity.
    \item We assess GPT-4o performance on failure explanation and categorization tasks, finding strong semantic alignment (Cosine = 0.93, GPTScore = 0.93) and 80\% exact-match accuracy, demonstrating that reduced logs preserve the information needed for downstream reasoning.
    \item We demonstrate that embedding-based classifiers can automate log-line relevance detection with near-human accuracy (97\%), enabling scalable, sustainable integration of semantics-aware filtering into CI workflows.
\end{itemize}

Section~\ref{sec:Related_work} reviews related work, Section~\ref{sec:StudyDesign} describes our study design, and Section~\ref{sec:RQs} presents our research questions and analyses. Section~\ref{sec:Discussion} discusses implications, Section~\ref{sec:Threats_to_validity} outlines threats to validity, and Section~\ref{sec:Conclusion} concludes.

\begin{figure*}[!t]
\includegraphics[width=\textwidth]{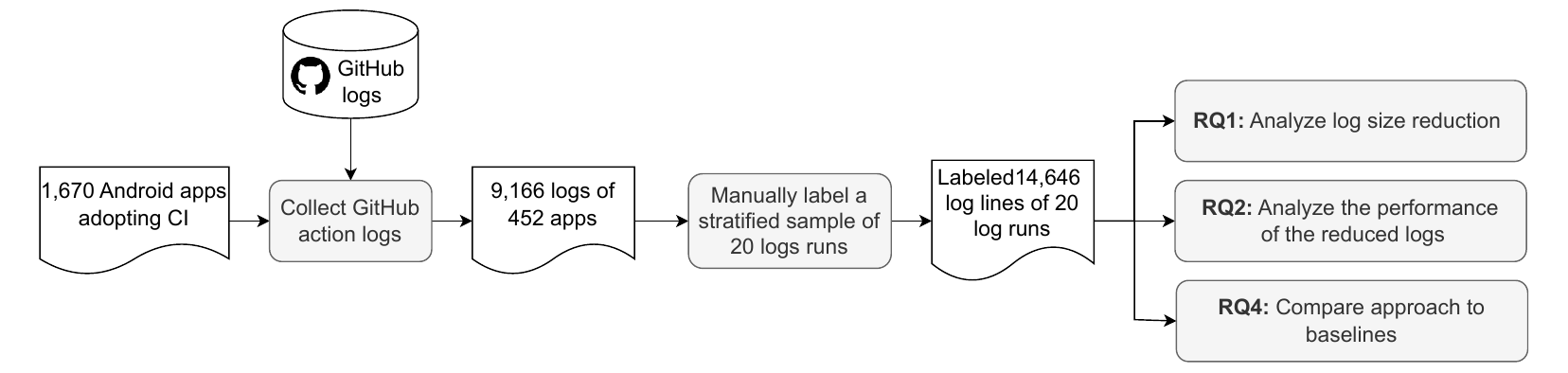}
\caption{An overview of LogSieve evaluation.}
\Description{Overview of the study methodology for RQ1, RQ2, and RQ4.}
\label{fig:data_collection}  
\end{figure*}

\section{Related Work}
\label{sec:Related_work}

\subsection{CI/CD Log Analysis and Mobile CI Context}
Logs play a central role in CI/CD workflows, enabling developers to diagnose failures, monitor pipelines, and assess build health. However, CI logs, especially in mobile ecosystems, tend to be long, unstructured, and noisy. This reflects the broader complexity and heterogeneity of CI workflows~\cite{abrokwah2025empirical}, which is also evident in Android projects~\cite{ghaleb2025ci} and can result in logs containing many low-information lines.
Prior studies of mobile applications have shown that Android apps often exhibit less-maintained and harder-to-parse logs~\cite{harty2021logging,zhu2019tools}, suggesting these tendencies extend to their CI workflows.
In root cause analysis for CI/CD workflows, logs are typically examined manually to find recurring patterns linked to different failure or error types~\cite{brandt2020logchunks,ghaleb2019noise}. This becomes increasingly challenging in mobile CI environments, where logs are large, noisy, and heterogeneous.
Despite these issues, the empirical analysis and reduction of CI logs remain relatively underexplored. Prior studies have examined how CI outcomes, such as build breakages, relate to pipeline characteristics and evolution, highlighting the importance of scalable diagnostic signals within CI workflows~\cite{ghaleb2022studying}.

\subsection{Classical Log Reduction and Compression}
Traditional log analysis has focused on redundancy removal and anomaly detection. Yao et al.~\cite{yao2021improving} improved compression methods for log management, while Xu et al.~\cite{xu2009detecting} applied template mining to detect large-scale system anomalies. Techniques such as \textit{LogZip}~\cite{zhu2019logzip} and
\textit{LogShrink}~\cite{li2024logshrink} aim to improve storage and indexing efficiency
but often retain task-irrelevant content. Likewise, Drain~\cite{he2017drain} and Spell~\cite{du2017spell} abstract logs structurally but remain agnostic to the \textit{semantic relevance} of individual lines for developer tasks. These compression and parsing techniques serve infrastructure needs (e.g., log indexing, deduplication) but are ill-suited to LLM-based reasoning, where unnecessary tokens inflate inference cost and obscure failure context.

\subsection{LLM-based Log Understanding}
Recent work has begun to explore large language models for log interpretation. Qi et al.~\cite{qi2023loggpt} introduced \textit{LogGPT}, a transformer-based approach for anomaly detection, while Huang et al.~\cite{huang2024lofi} proposed \textit{LoFI}, a prompt-driven method for extracting fault-indicating lines. However, these systems rely on LLM inference during the reduction process itself, increasing computational cost and limiting scalability. In contrast, \textit{LogSieve} performs pre-inference reduction, removing irrelevant lines before LLM reasoning while maintaining semantic fidelity.

\subsection{Sustainability and Software Engineering Implications}
Beyond efficiency, CI log verbosity also poses sustainability challenges. Each additional token processed by an LLM contributes to computational and energy overhead, compounding the environmental cost of model-assisted analysis~\cite{Koenig2025, Naumann2011, schwartz2020green}. By reducing token counts prior to inference, \textit{LogSieve} demonstrates how sustainability can be operationalized within software engineering workflows through lightweight, semantics-aware preprocessing.

\subsection{Summary and Research Gap}
Task-aware reduction has recently been explored in other LLM-based pipelines to control inference cost by prioritizing task-relevant inputs, though not in the context of CI logs or failure diagnosis~\cite{barnes2025task}. Prior work has primarily optimized logs for storage compression, structural abstraction, or inference-time filtering. While such methods improve infrastructure efficiency, they often neglect the \emph{semantic relevance} of log content for downstream developer tasks. \textit{LogSieve} addresses this gap by introducing RCA-aware, pre-inference reduction that explicitly models semantic relevance rather than syntactic redundancy.

Building on this distinction, our study explores four key research questions that evaluate \textit{LogSieve} from multiple perspectives:
(1) the potential for reducing CI log size while preserving diagnostic information (RQ1),
(2) the degree to which reduced logs retain semantic fidelity for LLM-based analysis (RQ2),
(3) the feasibility of automating relevance classification using embeddings and traditional ML models (RQ3), and
(4) the comparative advantages of RCA-aware filtering over structural or random baselines (RQ4).
Together, these questions bridge the gap between structure-first log management and semantics-first, sustainable CI analytics. Table~\ref{tab:technique-comparison} summarizes how \textit{LogSieve} compares to existing techniques.

\section{Study Design}
\label{sec:StudyDesign}

\begin{figure*}[!ht]
\includegraphics[width=1\textwidth]{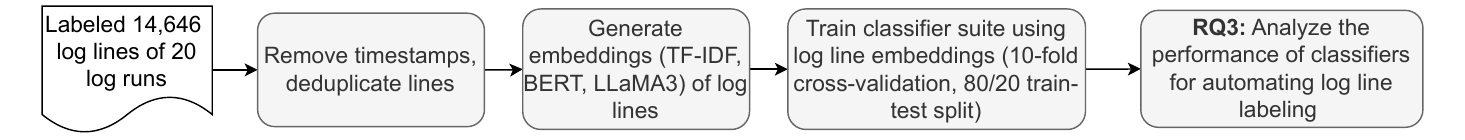}
\caption{An overview of automating LogSieve.}
\Description{Overview of the study methodology for RQ4.}
\label{fig:classifier_trianing}  
\end{figure*}

This study investigates how RCA-aware log reduction affects the accuracy and efficiency of LLM-based automation in Continuous Integration (CI) workflows. 
Figure~\ref{fig:data_collection} and Figure~\ref{fig:classifier_trianing} summarize the end-to-end workflow, from data collection and manual labeling to model training and evaluation. The design emphasizes empirical replicability, quantitative rigor, and fair comparison across reduction strategies. This staged structure mirrors how automation typically evolves in practice: manual annotation establishes a trusted ground truth, automated classification enables scalable deployment, and LLM-based evaluation verifies that the reduced logs preserve diagnostic fidelity.  By progressing from human judgment to machine learning and finally to LLM validation, each phase builds on the previous one to ensure that efficiency gains never compromise interpretability or practical relevance.

\subsection{Dataset Collection}
\label{subsec:dataset-collection}
Motivated by prior Android CI/CD studies~\cite{ghaleb2025ci}, we analyzed CI logs from 20 open-source Android projects that use GitHub Actions, selected to ensure diversity in workflow size, complexity, and logging patterns. To construct this corpus, we first collected 9{,}166 GitHub Actions workflow runs from 452 Android repositories; among these, 7{,}045 runs completed successfully, 1{,}806 failed, 219 were cancelled, and 96 were skipped. Because \textit{LogSieve} targets failure diagnostics, we focused on the failed runs. We drew a stratified random sample of 100 failed workflows to ensure coverage across project sizes, workflow configurations, and failure types (e.g., compilation, dependency, and environment errors). From this sample, we randomly selected 20 failed runs—one per repository—for detailed manual analysis, forming the annotated corpus used throughout RQ1--RQ4. Logs were collected via the GitHub REST~API as of March~11,~2025; given GitHub’s 90-day retention policy~\cite{GitHubDocs2025}, this dataset provides a timely snapshot of active CI usage. The dataset and supporting artifacts used in this study are publicly available as part of our replication package~\cite{barnes2026figshare}.

\subsection{Manual Annotation and Ground Truth Construction}
\label{subsec:annotation}
Each log line was manually labeled as RCA-relevant (\texttt{1}) or RCA-irrelevant (\texttt{0}) for understanding the cause of CI failure. Labeling followed a guideline adapted from prior CI/CD studies~\cite{ghaleb2025ci}, emphasizing the retention of diagnostic content (e.g., error traces, exception messages, task-level summaries) and removal of boilerplate or setup information (e.g., timestamps, dependency downloads, environment variables). Two co-authors independently labeled all 20 logs (14{,}646 lines in total). Inter-rater reliability was high (Cohen’s~$\kappa$~=~0.80)~\cite{cohen1960coefficient}. Disagreements were resolved in consensus meetings with a third co-author, yielding a validated ground truth dataset that serves as the foundation for automation experiments. In total, the annotators initially disagreed on 41 log lines. Representative contentious cases included whether lines related to accepting Android SDK licenses or specific \texttt{GITHUB\_TOKEN} permission settings should be retained as RCA-relevant in a CI context, as opposed to routine command or configuration details that were ultimately deemed irrelevant. Annotation was conducted independently by the two co-authors using a shared spreadsheet environment, ensuring that identical line ordering and contextual information were available during review. 
After labeling, we compared annotations programmatically to identify disagreements, which were then discussed in consensus meetings with the third author until full alignment was reached. This process not only standardized decisions on borderline cases but also reinforced inter-rater consistency, yielding a reproducible ground-truth dataset for downstream automation experiments.

The labeling guidelines were informed by the authors’ collective experience in software engineering and debugging.  When reviewing lines, we looked for cues indicating whether a line merely described routine processes or environment setup versus one that conveyed actionable diagnostic information such as error traces, exception messages, or test failures. 
Lines serving readability purposes only—such as progress indicators or decorative separators—were labeled as irrelevant, as they increase verbosity without aiding root-cause reasoning. 
These heuristics operationalize practical developer intuition about what constitutes meaningful versus superficial log content, anchoring the study’s focus on RCA-aware reduction for failure diagnosis.

\subsection{Model Training and Evaluation Setup}
\label{subsec:model-training}
To evaluate whether relevance classification can be automated at scale (RQ3), we trained a suite of supervised machine-learning classifiers using embeddings derived from three representation families. 
(1) \textbf{TF--IDF} captured lexical statistics and co-occurrence patterns, providing a strong and computationally inexpensive baseline for sparse text data. (2) \textbf{BERT} (\texttt{bert-base-uncased}) represented dense contextual semantics pre-trained on large natural language corpora, enabling sensitivity to surrounding log context such as error traces and stack frames.  (3) \textbf{LLaMA3} (\texttt{Meta-Llama-3-8B\-Instruct}) offered instruction-tuned embeddings that reflect how modern LLMs represent task prompts. 
Embeddings were precomputed and used as fixed feature vectors without fine-tuning, ensuring reproducibility and avoiding retraining overhead that would undermine \textit{LogSieve}'s lightweight and sustainable design goals.

The classifier suite spanned diverse learning paradigms: linear models (\textit{Logistic Regression}, \textit{Linear SVM}, \textit{SGD Logistic}); kernel and neural models (\textit{SVM RBF}, \textit{MLP}); ensemble methods (\textit{Random Forest}, \textit{XGBoost}, \textit{LightGBM}); and baselines (\textit{Nearest Centroid}, \textit{Dummy Classifier}).  This diversity balances interpretability and scalability: linear and tree models can be efficiently deployed in CI pipelines, while kernel and neural variants test the upper bound of achievable accuracy.  For models benefiting from normalization or dimensionality reduction, pipelines applied \textit{StandardScaler} and optionally \textit{PCA}, with component counts tuned via grid search over \{32, 64, 128, 256\}. These component sizes approximate the dimensionalities of typical log-parsing embeddings, ensuring comparability across model families.

Training used stratified 10-fold cross-validation with an 80/20 train--test split to preserve class balance between relevant and irrelevant lines. This procedure produced 10 train/test partitions (one per fold), minimizing sampling bias and providing stable estimates of generalization accuracy. We report accuracy, weighted~F1, precision, and recall on the held-out test set.

All models were implemented using scikit-learn~\cite{pedregosa2011scikit} and open-source gradient-boosting libraries XGBoost~\cite{chen2016xgboost} and LightGBM~\cite{ke2017lightgbm}.

\subsection{Evaluation Metrics and LLM Tasks}
\label{subsec:evaluation}
We computed four primary metrics to evaluate reduction performance and semantic preservation:

\begin{description}[leftmargin=0.5cm,labelsep=0.3cm]
  \item[\textbf{Line-level reduction ($Line\text{-}Red$)}:] 
  The percentage of log lines removed ($Removed\_Lines \times 100 / Total\_Lines$).

  \item[\textbf{Token-level reduction ($Token\text{-}Red$)}:] 
  The percentage of tokens removed ($Removed\_Tokens \times 100 / Total\_Tokens$), measured using OpenAI’s \texttt{tiktoken}\footnote{\url{https://github.com/openai/tiktoken}} tokenizer.

  \item[\textbf{Semantic similarity:}] 
  We compared GPT-4o responses generated from reduced versus full logs using cosine similarity~\cite{salton1983}, BERTScore~\cite{zhang2019bertscore}, ROUGE-1/L~\cite{lin-2004-rouge}, BLEU~\cite{papineni2002bleu}, and GPTScore~\cite{fu-etal-2024-gptscore}. These metrics collectively capture both semantic and lexical fidelity.

  \item[\textbf{Classification agreement:}] 
  Exact-match accuracy between full- and reduced-log predictions for the categorization task (Prompt~2), quantifying whether reduction alters the categorical outcome of LLM reasoning.
\end{description}

The following rationale clarifies why each metric was chosen and how they collectively capture both surface and semantic fidelity.

These metrics were selected to capture complementary aspects of information preservation.  Cosine similarity quantifies vector-level alignment between embeddings and reflects coarse semantic overlap, while BERTScore measures fine-grained contextual correspondence using token-level attention. 
ROUGE and BLEU evaluate lexical overlap and surface variability, ensuring that semantic alignment is not achieved by ignoring syntactic fidelity. 
GPTScore provides an LLM-judged perspective on response quality that better correlates with human preferences than purely statistical metrics. 
Together, these measures balance lexical and semantic dimensions of fidelity, offering a holistic assessment of how much diagnostic information is retained after reduction.

We selected GPT-4o for this study because it represented the state of the art among accessible large language models at the time of experimentation and was available via a stable API interface. Our goal was to assess whether the \textit{LogSieve} approach yields meaningful gains with a high-performing frontier model before extending evaluation to other architectures. In future work, we plan to replicate these experiments with both open-source and proprietary LLMs to determine whether \textit{LogSieve}’s benefits are range-bound across model families. To assess information preservation, we used two GPT-4o prompts:

\promptbox{Prompt 1 (Explanation):}{%
Here is the log output from a GitHub Action workflow run: [log.txt]. In at most 500 words, please explain why this workflow failed.
}

\promptbox{Prompt 2 (Categorization):}{%
Here is the log output from a GitHub Action workflow run: [log.txt]. Please provide a category for the run failure. Only provide the answer. Do not include any additional reasons or details.
}

Prompt~1 reflects GitHub’s “Explain Error” Copilot use case, while Prompt~2 captures coarse-grained failure categorization as used in CI dashboards. Model responses for full versus reduced logs were compared using the metrics above. The resulting measurements (Tables~\ref{tab:combined_reduction}–\ref{tab:llm_combined_eval}) form the basis for RQ1–RQ4.

\subsection{Baseline Comparison Design}
\label{subsec:baseline-design}
Finally, for RQ4, we benchmarked \textit{LogSieve} against two baselines: the compression-based technique \textit{LogZip}~\cite{zhu2019logzip} and a random-line removal control. Each baseline was applied to the same CI logs, and the resulting reduced logs were re-evaluated using the same GPT-4o prompts and similarity metrics. This design enables a fair comparison of \textit{LogSieve} against structural and non-semantic reductions in terms of both compression ratio and semantic fidelity.

\section{Research Questions}
\label{sec:RQs}
This study addresses four research questions designed to evaluate the effectiveness and scalability of \textit{LogSieve}.

\subsection{RQ1: How much can CI log size be reduced by removing RCA-irrelevant lines?}
\paragraph{Motivation.}
CI logs are often verbose and contain a mix of relevant and irrelevant information, particularly in failed workflows. Reducing this noise can lower processing costs and improve the focus of downstream tools. This RQ investigates the extent to which log size can be reduced by removing RCA-irrelevant lines (root cause analysis in this context), providing insight into the potential efficiency gains of lightweight pre-reduction.

\paragraph{Approach.}
We began with 9{,}166 GitHub Actions (GA) workflow runs collected from 452 unique Android repositories that adopted CI pipelines. Of these, 7{,}045 runs completed successfully, 1{,}806 failed, 219 were cancelled, and 96 were skipped. Because LogSieve targets failure-related diagnostics, we focused on failed runs only. From this subset, we created a stratified sample of 100 failed runs to ensure representation across diverse project sizes, workflow configurations, and failure types (e.g., compilation, dependency, and environment errors). We then randomly selected 20 runs from this stratified sample for detailed manual analysis, forming the dataset used for the root-cause analysis (RCA) and failure categorization tasks evaluated in this paper.

Each log line within these 20 runs was manually annotated as \textit{RCA-relevant}~(1) or \textit{RCA-irrelevant}~(0) for understanding the failure. Annotation guidelines were adapted from prior CI/CD empirical studies~\cite{ghaleb2025ci}, emphasizing the retention of diagnostic evidence (e.g., error traces, exception messages, task-level summaries) and the removal of routine or boilerplate information (e.g., timestamps, dependency downloads, environment setup). Two co-authors independently labeled all lines using these guidelines. Inter-rater reliability, computed via Cohen’s~$\kappa$, was~0.80, indicating strong agreement~\cite{cohen1960coefficient}. Disagreements were discussed in consensus meetings with a third co-author until full alignment was reached, ensuring a consistent ground truth. The finalized annotations were then used to measure both line-level and token-level reductions after removing RCA-irrelevant content.

\paragraph{Results.}
Table~\ref{tab:combined_reduction} summarizes both line- and token-level reductions across projects. On average, 42\% of log lines and 40\% of tokens were removed. Some logs were highly reducible (e.g., \texttt{marunjar\allowbreak/anewjku}, 85\% line reduction), while others retained most content (e.g., \texttt{cyb3rko\allowbreak/flashdim}, 11\%).

These results suggest that CI logs—especially those associated with failed workflows—contain substantial portions of non-essential lines. \textit{LogSieve} offers a generalizable method for reducing log verbosity prior to LLM-based processing, without requiring system-specific tuning.

\begin{rqbox}
\textbf{RQ1 summary:}
CI logs are highly redundant. On average, \textit{LogSieve} removed \textbf{42\% of lines}
and \textbf{40\% of tokens} while preserving key diagnostic context. Pre-inference RCA-aware filtering can reduce cost and proportionally cut energy use (and associated emissions) without sacrificing interpretability.
\end{rqbox}

\begin{table}[t]
\caption{Log and token reduction across Android repositories.}
\label{tab:combined_reduction}
\centering
\scriptsize
\setlength{\tabcolsep}{3.5pt}
\begin{tabular}{lrrrrrrr}
\toprule
\textbf{Repository} &
\makecell{\textbf{Total}\\\textbf{Lines}} &
\makecell{\textbf{Removed}\\\textbf{Lines}} &
\makecell{\textbf{Lines}\\\textbf{Kept}} &
\makecell{\textbf{Line}\\\textbf{Red.}} &
\makecell{\textbf{Full}\\\textbf{Tokens}} &
\makecell{\textbf{Tokens}\\\textbf{Kept}} &
\makecell{\textbf{Token}\\\textbf{Red.}} \\
\midrule
\texttt{rafayali/movies}                  & 277  & 109  & 168  & 39\% & 8{,}834  & 5{,}775  & 35\% \\
\makecell[l]{\texttt{mcastillof/}\\\texttt{FakeTraveler}} & 276  & 101  & 175  & 37\% & 8{,}372  & 6{,}025  & 28\% \\
\texttt{meditohq/medito}                  & 41   & 10   & 31   & 24\% & 1{,}090  & 813     & 25\% \\
\texttt{SecUSo/backup}                    & 2{,}052 & 1{,}253 & 799  & 61\% & 74{,}157 & 28{,}373 & 62\% \\
\texttt{fm-sys/snapdrop}                  & 268  & 85   & 183  & 32\% & 10{,}598 & 7{,}416  & 30\% \\
\texttt{alex/MonsterComp}                 & 188  & 37   & 151  & 20\% & 5{,}442  & 4{,}501  & 17\% \\
\makecell[l]{\texttt{thesandipv/}\\\texttt{watchdone}}    & 2{,}192 & 291  & 1{,}901 & 13\% & 67{,}613 & 60{,}155 & 17\% \\
\texttt{Vishnu/Quotes}                    & 866  & 489  & 377  & 56\% & 29{,}130 & 13{,}835 & 53\% \\
\texttt{cyb3rko/flashdim}                 & 386  & 43   & 343  & 11\% & 15{,}814 & 14{,}641 & 7\%  \\
\texttt{hide1202/MovieDB}                 & 1{,}400 & 603  & 797  & 43\% & 45{,}367 & 26{,}784 & 41\% \\
\texttt{marunjar/anewjku}                 & 522  & 444  & 78   & 85\% & 16{,}737 & 2{,}792  & 83\% \\
\texttt{CrazyM/ToDont}                    & 196  & 58   & 134  & 27\% & 6{,}459  & 5{,}091  & 21\% \\
\texttt{dashpay/wallet}                   & 721  & 518  & 203  & 72\% & 23{,}868 & 7{,}595  & 68\% \\
\texttt{Graphene/PdfView}                 & 401  & 133  & 268  & 33\% & 12{,}231 & 8{,}771  & 28\% \\
\texttt{aivan/kpassnotes}                 & 1{,}157 & 949  & 208  & 82\% & 36{,}626 & 7{,}100  & 81\% \\
\texttt{ygg/android}                      & 27   & 4    & 23   & 15\% & 1{,}069  & 792     & 26\% \\
\texttt{metabrainz/MBZ}                   & 200  & 79   & 121  & 40\% & 6{,}470  & 4{,}304  & 33\% \\
\makecell[l]{\texttt{lollipopkit/}\\\texttt{flutter}}     & 206  & 84   & 122  & 41\% & 6{,}955  & 4{,}316  & 38\% \\
\texttt{TrianguloY/URL}                   & 1{,}714 & 941  & 773  & 55\% & 100{,}344 & 29{,}723 & 70\% \\
\makecell[l]{\texttt{ofalvai/}\\\texttt{HabitBuilder}}    & 1{,}550 & 787  & 772  & 50\% & 53{,}681 & 28{,}296 & 47\% \\
\addlinespace
\midrule
\textbf{Average}                           & \textbf{732} & \textbf{350} & \textbf{381} & \textbf{42\%} & \textbf{26{,}543} & \textbf{13{,}355} & \textbf{40\%} \\
\bottomrule
\end{tabular}
\end{table}

\subsection{RQ2: To what extent do reduced logs preserve semantic information necessary for LLM-based root cause analysis?}

\paragraph{Motivation.}
LLMs are increasingly used for root cause analysis and failure triage in CI and cloud systems. Recent studies have shown that LLM-based agents can match or outperform traditional methods in diagnosing incidents~\cite{roy2024exploring,wang2024rcagent}. However, these approaches typically rely on full log data. It remains unclear whether reduced logs preserve enough diagnostic information to support accurate LLM-driven analysis.

\paragraph{Approach.}
We evaluated GPT‑4o on two CI-related tasks: failure explanation and failure classification. As detailed in Section~\ref{sec:StudyDesign}, we used two prompts and applied them to both full and reduced logs: \textit{Prompt 1} for generating explanations, and \textit{Prompt 2} for classifying failure types. We assessed \textit{Prompt 1} using semantic similarity metrics (Cosine Similarity, BERTScore, GPTScore) and lexical metrics (ROUGE-1/L, BLEU). We used exact match accuracy for \textit{Prompt 2}. This setup quantifies whether reduced logs retain the context necessary for effective LLM responses.
\begin{table}[tb]
\caption{LLM Response Evaluation Across Repositories: Semantic Similarity and Classification Accuracy}
\label{tab:llm_combined_eval}
\centering
\scriptsize
\setlength{\tabcolsep}{4pt}
\begin{tabular}{lrrrrrrc}
\toprule
\textbf{Repository} & \textbf{CosSim} & \textbf{BERT-F1} & \textbf{R1-F1} & \textbf{RL-F1} & \textbf{BLEU} & \makecell{\textbf{GPT}\\\textbf{Score}} & \makecell{\textbf{Exact}\\\textbf{Match}} \\
\midrule
\texttt{rafayali/movies}                & 0.98 & 0.85 & 0.82 & 0.66 & 0.53 & 1.00 & 1 \\
\makecell[l]{\texttt{mcastillof/}\\\texttt{FakeTraveler}} & 0.92 & 0.77 & 0.72 & 0.34 & 0.34 & 0.90 & 1 \\
\texttt{meditohq/medito}                & 0.94 & 0.76 & 0.73 & 0.49 & 0.38 & 0.95 & 1 \\
\texttt{SecUSo/backup}                  & 0.84 & 0.74 & 0.68 & 0.49 & 0.40 & 0.90 & 0 \\
\texttt{fm-sys/snapdrop}                & 0.94 & 0.77 & 0.68 & 0.32 & 0.32 & 0.90 & 1 \\
\texttt{alex/MonsterComp}               & 0.94 & 0.80 & 0.74 & 0.45 & 0.41 & 0.90 & 0 \\
\makecell[l]{\texttt{thesandipv/}\\\texttt{watchdone}}    & 0.97 & 0.82 & 0.78 & 0.57 & 0.51 & 0.95 & 1 \\
\texttt{Vishnu/Quotes}                  & 0.95 & 0.82 & 0.79 & 0.56 & 0.49 & 0.95 & 1 \\
\texttt{cyb3rko/flashdim}               & 0.96 & 0.82 & 0.73 & 0.57 & 0.54 & 0.90 & 0 \\
\texttt{hide1202/MovieDB}               & 0.94 & 0.85 & 0.79 & 0.56 & 0.56 & 0.95 & 1 \\
\texttt{marunjar/anewjku}               & 0.86 & 0.72 & 0.62 & 0.29 & 0.28 & 0.80 & 0 \\
\texttt{CrazyM/ToDont}                  & 0.96 & 0.78 & 0.74 & 0.45 & 0.38 & 1.00 & 1 \\
\texttt{dashpay/wallet}                 & 0.97 & 0.85 & 0.79 & 0.62 & 0.51 & 0.95 & 1 \\
\texttt{Graphene/PdfView}               & 0.91 & 0.76 & 0.65 & 0.37 & 0.32 & 0.90 & 1 \\
\texttt{aivan/kpassnotes}               & 0.92 & 0.74 & 0.65 & 0.30 & 0.23 & 0.90 & 1 \\
\texttt{ygg/android}                    & 0.96 & 0.79 & 0.76 & 0.50 & 0.46 & 1.00 & 1 \\
\makecell[l]{\texttt{metabrainz}/\\\texttt{MBZ}}          & 0.85 & 0.78 & 0.70 & 0.44 & 0.37 & 0.90 & 1 \\
\makecell[l]{\texttt{lollipopkit}/\\\texttt{flutter}}  & 0.95 & 0.79 & 0.74 & 0.50 & 0.50 & 0.95 & 1 \\
\makecell[l]{\texttt{TrianguloY}/\\\texttt{URL}}          & 0.95 & 0.78 & 0.71 & 0.43 & 0.37 & 0.90 & 1 \\
\makecell[l]{\texttt{ofalvai}/\\\texttt{HabitBuilder}}    & 0.93 & 0.81 & 0.75 & 0.44 & 0.44 & 0.90 & 1 \\
\addlinespace
\cmidrule(lr){1-8}
\textbf{Min}    & \textbf{0.84} & \textbf{0.72} & \textbf{0.62} & \textbf{0.29} & \textbf{0.23} & \textbf{0.80} & \multirow{4}{*}{\makecell{\textbf{Total}\\\textbf{16 / 20}\\\textbf{(80\%)}}} \\
\textbf{Mean}   & \textbf{0.93} & \textbf{0.79} & \textbf{0.73} & \textbf{0.47} & \textbf{0.42} & \textbf{0.93} &  \\
\textbf{Median} & \textbf{0.94} & \textbf{0.78} & \textbf{0.73} & \textbf{0.47} & \textbf{0.41} & \textbf{0.90} &  \\
\textbf{Max}    & \textbf{0.98} & \textbf{0.85} & \textbf{0.82} & \textbf{0.66} & \textbf{0.56} & \textbf{1.00} &  \\
\bottomrule
\end{tabular}
\end{table}
\paragraph{Results.}
Table~\ref{tab:llm_combined_eval} presents semantic similarity scores between responses generated from full and reduced logs. Cosine Similarity averaged 0.93, BERTScore (F1) was 0.79, and the GPTScore—derived from a secondary GPT judgment—averaged 0.93. While lexical metrics such as ROUGE-L (0.47) and BLEU (0.42) were more sensitive to surface variation, overall semantic alignment remained strong.

For Prompt 2, the last column of Table~\ref{tab:llm_combined_eval} shows that 16 out of 20 reduced-log responses matched the full-log classification exactly (80\% accuracy). This result is notable given an average token reduction of 40\%, demonstrating that \textit{LogSieve} preserves key diagnostic information required for LLM-driven analysis.

These findings suggest that \textit{LogSieve} enables meaningful reductions in log size without sacrificing the quality of LLM responses for CI failure diagnosis.

\begin{rqbox}
\textbf{RQ2 summary:}
Despite a 40\% token reduction, GPT-4o responses maintained high fidelity (\textbf{CosSim = 0.93}, \textbf{GPTScore = 0.93}, and \textbf{80\% exact-match accuracy}).  RCA-aware reduction preserves essential semantics, enabling efficient and accurate
LLM-based CI log analysis.
\end{rqbox}

\begin{table}[t]
\caption{Performance of ML classifiers across embeddings for automated RCA-relevance classification. Bold values indicate the best-performing model within each embedding family.}
\label{tab:embeddings-full}
\centering
\scriptsize
\begin{tabular}{llcccc}
\toprule
\textbf{Embedding} & \textbf{Model} & \textbf{Accuracy} & \textbf{F1 (w)} & \textbf{Precision (w)} & \textbf{Recall (w)} \\
\midrule
\multirow{10}{*}{TF--IDF}
 & Logistic Regression (L2) & \textbf{0.971} & \textbf{0.971} & \textbf{0.971} & \textbf{0.971} \\
 & MLP (small)              & 0.968 & 0.968 & 0.968 & 0.968 \\
 & XGBoost                  & 0.967 & 0.967 & 0.968 & 0.967 \\
 & LightGBM                 & 0.967 & 0.967 & 0.967 & 0.967 \\
 & Random Forest            & 0.967 & 0.967 & 0.967 & 0.967 \\
 & SVM RBF                  & 0.966 & 0.966 & 0.966 & 0.966 \\
 & Nearest Centroid         & 0.963 & 0.963 & 0.963 & 0.963 \\
 & Linear SVM               & 0.958 & 0.958 & 0.959 & 0.958 \\
 & SGD Logistic             & 0.944 & 0.944 & 0.945 & 0.944 \\
 & Dummy (stratified)       & 0.504 & 0.504 & 0.504 & 0.504 \\
\midrule
\multirow{10}{*}{BERT}
 & Logistic Regression (L2) & \textbf{0.971} & \textbf{0.971} & \textbf{0.971} & \textbf{0.971} \\
 & MLP (small)              & 0.968 & 0.968 & 0.968 & 0.968 \\
 & XGBoost                  & 0.967 & 0.967 & 0.968 & 0.967 \\
 & LightGBM                 & 0.967 & 0.967 & 0.967 & 0.967 \\
 & Random Forest            & 0.967 & 0.967 & 0.967 & 0.967 \\
 & SVM RBF                  & 0.966 & 0.966 & 0.966 & 0.966 \\
 & Nearest Centroid         & 0.963 & 0.963 & 0.963 & 0.963 \\
 & Linear SVM               & 0.958 & 0.958 & 0.959 & 0.958 \\
 & SGD Logistic             & 0.944 & 0.944 & 0.945 & 0.944 \\
 & Dummy (stratified)       & 0.504 & 0.504 & 0.504 & 0.504 \\
\midrule
\multirow{10}{*}{LLaMA3}
 & SVM RBF                  & \textbf{0.972} & \textbf{0.972} & \textbf{0.973} & \textbf{0.972} \\
 & SGD Logistic             & 0.972 & 0.972 & 0.973 & 0.972 \\
 & MLP (small)              & 0.971 & 0.971 & 0.971 & 0.971 \\
 & LightGBM                 & 0.971 & 0.971 & 0.971 & 0.971 \\
 & XGBoost                  & 0.971 & 0.971 & 0.971 & 0.971 \\
 & Logistic Regression (L2) & 0.969 & 0.969 & 0.969 & 0.969 \\
 & Linear SVM               & 0.969 & 0.969 & 0.969 & 0.969 \\
 & Random Forest            & 0.966 & 0.966 & 0.967 & 0.966 \\
 & Nearest Centroid         & 0.854 & 0.855 & 0.855 & 0.854 \\
 & Dummy (stratified)       & 0.500 & 0.500 & 0.500 & 0.500 \\
\bottomrule
\end{tabular}
\end{table}
\subsection{RQ3: How effective are machine learning models using different embeddings at automating log reduction?}

\paragraph{Motivation.}
Building on the manually labeled ground truth from RQ1 and RQ2, we next investigated whether embeddings could automate the relevance classification process. While manual labeling is useful for pilot evaluation, it is not feasible in practice. To make \textit{LogSieve} applicable to real-world CI workflows, relevance classification must be automated. Embedding-based models provide a natural mechanism for this task by capturing semantic features of log lines and supporting binary classification into RCA-relevant and RCA-irrelevant categories.

\paragraph{Approach.}
We evaluated three embedding families to determine their effectiveness for automated log reduction:
\begin{itemize}
    \item \textbf{TF--IDF}: a sparse lexical representation capturing term frequency and inverse document frequency,
    \item \textbf{BERT (bert-base-uncased)}: dense contextual embeddings pre-trained on large text corpora, and
    \item \textbf{LLaMA3 (Meta-Llama-3-8B-Instruct)}: instruction-tuned embeddings representing modern large language model features.
\end{itemize}

For each embedding type, we trained ten supervised classifiers spanning linear, kernel, neural, and ensemble families: Logistic Regression (L2), Linear SVM, SGD Logistic, SVM with RBF kernel, a small MLP, Random Forest, XGBoost, LightGBM, Nearest Centroid, and a Dummy (stratified) baseline. Embeddings were treated as fixed feature vectors (not fine-tuned), and all models were trained under identical 10-fold stratified cross-validation. Hyperparameters for PCA-based models were tuned over \{32, 64, 128, 256\} components using grid search. Model performance was evaluated on a held-out 20\% test set using accuracy, weighted F1, precision, and recall.

\paragraph{Results.}
Table~\ref{tab:embeddings-full} presents the performance of all classifiers across the three embedding families. All embeddings produced strong results, confirming that both lexical and contextual representations can support accurate automation of log-line relevance classification.

For \textbf{TF--IDF} embeddings, \textit{Logistic Regression (L2)} achieved the best accuracy of \textbf{0.97}, with tree- and boosting-based models (\textit{Random Forest}, \textit{XGBoost}, \textit{LightGBM}) and the \textit{SVM (RBF)} kernel following closely ($\geq$\,0.96). This shows that sparse lexical features already capture much of the discriminative signal between relevant and irrelevant lines, such as the co-occurrence of failure keywords, file-path patterns, or stack-trace indicators.  

For \textbf{BERT} embeddings, \textit{Logistic Regression (L2)} again obtained the highest accuracy (\textbf{0.97}), and most other models performed within a small margin. The similarity of BERT and TF--IDF scores suggests that contextual embeddings provide only marginal gains for this binary task, likely because CI-log language and structure are relatively stable across projects and contain limited ambiguity compared with natural text.

For \textbf{LLaMA3} embeddings, the best result was produced by \textit{SVM (RBF)} (\textbf{0.97}), followed closely by \textit{SGD Logistic} and \textit{MLP} (0.97 each). Although LLaMA3 embeddings are high-dimensional and instruction-tuned, they did not yield substantial improvements over simpler representations for this classification setting. Nevertheless, their comparable performance confirms that \textit{LogSieve} can integrate seamlessly with instruction-tuned embedding services if such representations are already available within an organization’s LLM infrastructure.

Across all embeddings, weighted~F1, precision, and recall remained consistently high ($\approx$\,0.96--0.97), indicating balanced performance and minimal class bias.

Overall, these results demonstrate that classical machine-learning classifiers can emulate human relevance judgments with near-human accuracy and low computational cost, providing a practical path for scaling \textit{LogSieve} to large industrial CI environments without invoking expensive LLM inference.

\begin{rqbox}
\textbf{RQ3 summary:}
Classical machine-learning classifiers, such as \textit{Logistic Regression} and \textit{SVM (RBF)}, achieve near-human labeling accuracy across all embedding types. These results demonstrate that \textit{LogSieve} can reliably automate RCA-relevance detection at scale within CI pipelines.
\end{rqbox}
\begin{table}[t]
\caption{RQ4 comparison of \textit{LogSieve}, \textit{LogZip}, and \textit{Random} baselines against full logs. Means across 20 repositories (macro-averages).}
\label{tab:rq4_baselines}
\centering
\setlength{\tabcolsep}{3.5pt}
\begin{tabular}{lcccccc}
\toprule
\multicolumn{7}{c}{\textbf{Prompt~1 (Failure Explanation): mean similarity to Full}} \\
\addlinespace[1pt]
\midrule
 & CosSim & BERT-F1 & R1-F1 & RL-F1 & BLEU & GPTScore \\
\addlinespace[1pt]
\midrule
\textit{LogSieve} & \textbf{0.93} & \textbf{0.79} & \textbf{0.73} & \textbf{0.47} & \textbf{0.42} & \textbf{0.93} \\
LogZip            & 0.70 & 0.68 & 0.51 & 0.26 & 0.09 & 0.41 \\
Random            & 0.90 & 0.76 & 0.66 & 0.38 & 0.23 & 0.86 \\
\addlinespace[1pt]
\midrule
\multicolumn{7}{c}{\textbf{Prompt~2 (Failure Categorization): mean scores vs Full}} \\
\addlinespace[1pt]
\midrule
 & CosSim & BERT-F1 & R1-F1 & RL-F1 & BLEU & GPTScore \\
\addlinespace[1pt]
\midrule
\textit{LogSieve} & \textbf{0.92} & \textbf{0.93} & \textbf{0.87} & \textbf{0.87} & \textbf{0.29} & \textbf{0.94} \\
LogZip            & 0.51 & 0.67 & 0.23 & 0.23 & 0.07 & 0.47 \\
Random            & 0.87 & 0.89 & 0.77 & 0.77 & 0.29 & 0.89 \\
\addlinespace[1pt]
\midrule
\multicolumn{7}{c}{\textbf{Prompt~2: Exact-match Accuracy (Full vs.\ Reduced)}} \\
\midrule
\textit{LogSieve} & \multicolumn{6}{c}{0.80} \\
LogZip            & \multicolumn{6}{c}{0.20} \\
Random            & \multicolumn{6}{c}{0.70} \\
\bottomrule
\end{tabular}
\end{table}

\subsection{RQ4: How does \textit{LogSieve} compare to baseline approaches?}
\paragraph{Motivation.} While RQ1–RQ3 examined \textit{LogSieve}'s intrinsic reduction and automation performance, this question contextualizes its effectiveness against two alternative baselines:
\begin{itemize}
    \item \textbf{LogZip}. A widely used compression-based technique that clusters repetitive log templates to reduce redundancy without considering RCA-relevance~\cite{zhu2019logzip}. This baseline tests whether structural compression alone can preserve the information required for LLM-based analysis.
    \item \textbf{Random-line removal}. A sanity-check baseline that randomly removes lines to match \textit{LogSieve}'s average 42\% reduction rate. Following standard practice in model evaluation~\cite{megahed2024comparing}, this ensures that improvements are due to semantic relevance rather than reduction volume.
\end{itemize}
\paragraph{Approach.} For each repository, we produced three reduced versions of the logs:  
(i) \textit{LogSieve}-reduced (RCA-aware filtering),  
(ii) \textit{LogZip}-reduced (template-based compression), and  
(iii) \textit{Random}-reduced (random line removal).  
We re-evaluated these logs on the same GPT-4o downstream tasks from RQ2—\emph{Prompt~1} (failure explanation) and \emph{Prompt~2} (failure categorization)—using identical metrics: Cosine Similarity (CosSim), BERTScore~F1, ROUGE-1/L~F1, BLEU, GPTScore, and exact-match accuracy (Prompt~2 only).  We report the mean values across the 20 repositories to compare semantic and categorical fidelity under equivalent reduction ratios.
\paragraph{Results.} Table~\ref{tab:rq4_baselines} compares the three reduction strategies. Across both tasks, \textit{LogSieve} consistently outperformed the structural (\textit{LogZip}) and random baselines.   For \emph{Prompt~1} (failure explanation), \textit{LogSieve} achieved the highest semantic alignment with full logs (CosSim~=~0.93, GPTScore~=~0.93), compared to \textit{LogZip} (CosSim~=~0.70) and \textit{Random} (CosSim~=~0.90). For \emph{Prompt~2} (failure categorization), \textit{LogSieve} preserved categorical accuracy (exact-match~=~0.80), compared to 0.20 for \textit{LogZip} and 0.70 for \textit{Random}. Although the random baseline occasionally retained surface similarity, it degraded categorical fidelity and interpretability, confirming that \textit{LogSieve}'s RCA-aware filtering—rather than reduction magnitude—drives superior preservation of LLM reasoning fidelity. Table~\ref{tab:rq4_baselines} reports macro-averaged performance across repositories; per-repository breakdowns for the baselines appear in Tables~\ref{tab:llm_logzip_eval} and~\ref{tab:llm_random_eval}.

To further understand variation across projects, we analyzed per-repository results for the baseline methods. As shown in Tables~\ref{tab:llm_logzip_eval} and~\ref{tab:llm_random_eval}, \textit{LogZip}’s performance varies widely across repositories (CosSim = 0.54--0.95; EM = 20\%), reflecting sensitivity to template diversity, whereas \textit{Random} reduction remains comparatively stable (CosSim $\approx$ 0.90; EM = 70\%). This dispersion underscores the advantage of \textit{LogSieve}’s RCA-aware filtering, which maintains both high semantic similarity and consistent classification fidelity across heterogeneous projects.

\begin{table}[t]
\caption{LLM response evaluation by repository for \textit{LogZip}: semantic similarity to Full (Prompt~1) and exact-match classification agreement (Prompt~2).}
\label{tab:llm_logzip_eval}
\centering
\scriptsize
\setlength{\tabcolsep}{4pt}
\begin{tabular}{lrrrrrrc}
\toprule
\textbf{Repository} & \textbf{CosSim} & \textbf{BERT-F1} & \textbf{R1-F1} & \textbf{RL-F1} & \textbf{BLEU} & \makecell{\textbf{GPT}\\\textbf{Score}} & \makecell{\textbf{Exact}\\\textbf{Match}} \\
\midrule
\texttt{rafayali/movies}                & 0.56 & 0.63 & 0.35 & 0.19 & 0.01 & 0.00 & 0 \\
\makecell[l]{\texttt{mcastillof/}\\\texttt{FakeTraveler}} & 0.88 & 0.75 & 0.69 & 0.34 & 0.21 & 0.90 & 1 \\
\texttt{meditohq/medito}                & 0.63 & 0.70 & 0.55 & 0.27 & 0.11 & 0.20 & 0 \\
\texttt{SecUSo/backup}                  & 0.91 & 0.71 & 0.66 & 0.43 & 0.26 & 0.90 & 1 \\
\texttt{fm-sys/snapdrop}                & 0.60 & 0.66 & 0.41 & 0.17 & 0.02 & 0.30 & 0 \\
\texttt{alex/MonsterComp}               & 0.94 & 0.79 & 0.73 & 0.48 & 0.28 & 0.90 & 1 \\
\makecell[l]{\texttt{thesandipv/}\\\texttt{watchdone}}    & 0.68 & 0.69 & 0.54 & 0.24 & 0.11 & 0.00 & 0 \\
\texttt{Vishnu/Quotes}                  & 0.71 & 0.66 & 0.51 & 0.24 & 0.07 & 0.60 & 0 \\
\texttt{cyb3rko/flashdim}               & 0.86 & 0.72 & 0.54 & 0.31 & 0.12 & 0.90 & 0 \\
\texttt{hide1202/MovieDB}               & 0.66 & 0.65 & 0.45 & 0.19 & 0.01 & 0.30 & 0 \\
\texttt{marunjar/anewjku}               & 0.84 & 0.73 & 0.63 & 0.32 & 0.12 & 0.90 & 0 \\
\texttt{CrazyM/ToDont}                  & 0.59 & 0.66 & 0.47 & 0.21 & 0.02 & 0.20 & 0 \\
\texttt{dashpay/wallet}                 & 0.56 & 0.61 & 0.45 & 0.18 & 0.01 & 0.20 & 0 \\
\texttt{Graphene/PdfView}               & 0.62 & 0.64 & 0.44 & 0.18 & 0.04 & 0.00 & 0 \\
\texttt{aivan/kpassnotes}               & 0.95 & 0.79 & 0.72 & 0.46 & 0.34 & 0.90 & 1 \\
\texttt{ygg/android}                    & 0.54 & 0.64 & 0.44 & 0.20 & 0.05 & 0.00 & 0 \\
\makecell[l]{\texttt{metabrainz}/\\\texttt{MBZ}}          & 0.62 & 0.64 & 0.43 & 0.20 & 0.01 & 0.20 & 0 \\
\makecell[l]{\texttt{lollipopkit}/\\\texttt{flutter}}  & 0.56 & 0.64 & 0.45 & 0.19 & 0.04 & 0.20 & 0 \\
\makecell[l]{\texttt{TrianguloY}/\\\texttt{URL}}          & 0.78 & 0.65 & 0.46 & 0.22 & 0.02 & 0.30 & 0 \\
\makecell[l]{\texttt{ofalvai}/\\\texttt{HabitBuilder}}    & 0.56 & 0.61 & 0.39 & 0.18 & 0.01 & 0.20 & 0 \\
\addlinespace
\cmidrule(lr){1-8}
\textbf{Min}    & \textbf{0.54} & \textbf{0.61} & \textbf{0.35} & \textbf{0.17} & \textbf{0.01} & \textbf{0.00} & \multirow{4}{*}{\makecell{\textbf{Total}\\\textbf{4 / 20}\\\textbf{(20\%)}}} \\
\textbf{Mean}   & \textbf{0.70} & \textbf{0.68} & \textbf{0.51} & \textbf{0.26} & \textbf{0.09} & \textbf{0.41} &  \\
\textbf{Median} & \textbf{0.65} & \textbf{0.66} & \textbf{0.46} & \textbf{0.21} & \textbf{0.05} & \textbf{0.25} &  \\
\textbf{Max}    & \textbf{0.95} & \textbf{0.79} & \textbf{0.73} & \textbf{0.48} & \textbf{0.34} & \textbf{0.90} &  \\
\bottomrule
\end{tabular}
\end{table}

\begin{table}[t]
\caption{LLM response evaluation by repository for \textit{Random} reduction: semantic similarity to Full (Prompt~1) and exact-match classification agreement (Prompt~2).}
\label{tab:llm_random_eval}
\centering
\scriptsize
\setlength{\tabcolsep}{4pt}
\begin{tabular}{lrrrrrrc}
\toprule
\textbf{Repository} & \textbf{CosSim} & \textbf{BERT-F1} & \textbf{R1-F1} & \textbf{RL-F1} & \textbf{BLEU} & \makecell{\textbf{GPT}\\\textbf{Score}} & \makecell{\textbf{Exact}\\\textbf{Match}} \\
\midrule
\texttt{rafayali/movies}                & 0.92 & 0.78 & 0.70 & 0.41 & 0.27 & 0.90 & 0 \\
\makecell[l]{\texttt{mcastillof/}\\\texttt{FakeTraveler}} & 0.92 & 0.80 & 0.74 & 0.41 & 0.36 & 0.90 & 1 \\
\texttt{meditohq/medito}                & 0.93 & 0.80 & 0.76 & 0.52 & 0.38 & 1.00 & 1 \\
\texttt{SecUSo/backup}                  & 0.93 & 0.76 & 0.67 & 0.42 & 0.27 & 1.00 & 0 \\
\texttt{fm-sys/snapdrop}                & 0.91 & 0.75 & 0.66 & 0.30 & 0.18 & 0.90 & 1 \\
\texttt{alex/MonsterComp}               & 0.93 & 0.77 & 0.68 & 0.43 & 0.28 & 0.90 & 1 \\
\makecell[l]{\texttt{thesandipv/}\\\texttt{watchdone}}    & 0.90 & 0.75 & 0.69 & 0.35 & 0.22 & 0.90 & 1 \\
\texttt{Vishnu/Quotes}                  & 0.88 & 0.72 & 0.63 & 0.29 & 0.15 & 0.90 & 1 \\
\texttt{cyb3rko/flashdim}               & 0.91 & 0.75 & 0.60 & 0.36 & 0.22 & 0.90 & 1 \\
\texttt{hide1202/MovieDB}               & 0.93 & 0.80 & 0.66 & 0.39 & 0.23 & 0.90 & 1 \\
\texttt{marunjar/anewjku}               & 0.91 & 0.75 & 0.65 & 0.36 & 0.20 & 0.90 & 1 \\
\texttt{CrazyM/ToDont}                  & 0.88 & 0.79 & 0.69 & 0.37 & 0.21 & 0.80 & 0 \\
\texttt{dashpay/wallet}                 & 0.95 & 0.76 & 0.68 & 0.37 & 0.15 & 0.90 & 1 \\
\texttt{Graphene/PdfView}               & 0.90 & 0.78 & 0.70 & 0.39 & 0.26 & 0.90 & 1 \\
\texttt{aivan/kpassnotes}               & 0.93 & 0.74 & 0.62 & 0.37 & 0.20 & 0.90 & 0 \\
\texttt{ygg/android}                    & 0.58 & 0.64 & 0.44 & 0.20 & 0.03 & 0.00 & 0 \\
\makecell[l]{\texttt{metabrainz}/\\\texttt{MBZ}}          & 0.90 & 0.78 & 0.70 & 0.39 & 0.26 & 0.90 & 1 \\
\makecell[l]{\texttt{lollipopkit}/\\\texttt{flutter}}  & 0.91 & 0.77 & 0.67 & 0.41 & 0.27 & 0.90 & 1 \\
\makecell[l]{\texttt{TrianguloY}/\\\texttt{URL}}          & 0.94 & 0.76 & 0.68 & 0.45 & 0.27 & 0.90 & 1 \\
\makecell[l]{\texttt{ofalvai}/\\\texttt{HabitBuilder}}    & 0.92 & 0.79 & 0.70 & 0.47 & 0.26 & 0.90 & 0 \\
\addlinespace
\cmidrule(lr){1-8}
\textbf{Min}    & \textbf{0.58} & \textbf{0.64} & \textbf{0.44} & \textbf{0.20} & \textbf{0.03} & \textbf{0.00} & \multirow{4}{*}{\makecell{\textbf{Total}\\\textbf{14 / 20}\\\textbf{(70\%)}}} \\
\textbf{Mean}   & \textbf{0.90} & \textbf{0.76} & \textbf{0.66} & \textbf{0.38} & \textbf{0.23} & \textbf{0.86} &  \\
\textbf{Median} & \textbf{0.92} & \textbf{0.76} & \textbf{0.67} & \textbf{0.38} & \textbf{0.24} & \textbf{0.90} &  \\
\textbf{Max}    & \textbf{0.95} & \textbf{0.80} & \textbf{0.76} & \textbf{0.52} & \textbf{0.38} & \textbf{1.00} &  \\
\bottomrule
\end{tabular}
\end{table}

\begin{rqbox}
\textbf{RQ4 summary:}
The baseline comparison confirms that \textit{LogSieve} achieves much higher semantic and categorical fidelity than both structural and random reductions at equivalent reduction ratios.  While \textit{LogZip} effectively compresses redundant patterns, it retains RCA-irrelevant lines that reduce interpretability. The Random baseline shows that volume reduction without modeling semantic relevance yields lower categorical fidelity and interpretability. Thus, \textit{LogSieve}'s RCA-aware filtering, not reduction magnitude, drives superior preservation of LLM reasoning fidelity.
\end{rqbox}

\section{Discussion}
\label{sec:Discussion}

This section interprets our findings across RQ1--RQ4 and discusses their implications for automated, sustainable CI analytics. Together, the results show that RCA-aware log reduction can substantially decrease inference cost while preserving diagnostic fidelity.

\subsection{Automation Feasibility}
Our automation experiments (RQ3) show that log-line relevance can be reliably inferred using embedding-based classifiers, suggesting that manual labeling can be eliminated for most CI contexts. Embedding features capture both syntax and semantics of log text, achieving up to 97\% accuracy across models. These results highlight the practicality of integrating \textit{LogSieve} as a continuously learning pre-filter in CI systems, where embeddings can be updated incrementally as logs evolve.

In practice, \textit{LogSieve} can operate as an intermediate step between log collection and artifact storage, automatically tagging or filtering lines with confidence scores produced by the relevance classifier. 
A configurable threshold determines whether a line is retained or suppressed, allowing practitioners to balance interpretability and compression according to project risk or compliance needs.  Low-confidence cases can be routed to full logs or developer review, providing a human-in-the-loop safeguard that maintains transparency during adoption.  This flexible integration model enables gradual rollout in production pipelines without disrupting existing CI infrastructure.

To sustain performance over time, retraining can occur periodically or be triggered when classifier confidence drops below a set threshold, signaling drift in workflow or log structure.  Because training and inference are lightweight, retraining can run offline or on dedicated CI runners with minimal overhead.  Together, these features position \textit{LogSieve} as a practical, low-maintenance automation component that reduces human effort while preserving the interpretability and auditability essential for modern software delivery.

\subsection{Effectiveness Depends on Log Quality}
Aggressive reduction, when guided by RCA-relevance, lowers input size without sacrificing LLM effectiveness. Most classification mismatches in Table~\ref{tab:llm_combined_eval} stemmed from vague or underspecified logs rather than information loss. This suggests that the quality and structure of the source logs fundamentally bound the achievable fidelity of reduction. Logs that contain clear stack traces, compiler diagnostics, or structured error messages allow both humans and models to identify relevant lines reliably, whereas loosely formatted or incomplete logs make relevance prediction inherently ambiguous. Future CI pipelines could therefore benefit from adopting structured logging conventions or schema-based instrumentation to maximize the downstream utility of semantics-aware reduction.

Building on these observations, \textit{LogSieve} can also be paired with lightweight domain heuristics (e.g., compiler or toolchain detectors) to recover task context that is missing or weakly expressed in unstructured logs. 
Integrating such heuristics into preprocessing would ensure that key diagnostic tokens are never removed while still allowing aggressive reduction of routine setup and environment information. This trade-off between log quality and reduction aggressiveness offers a promising direction for practical deployment, where developers can tune thresholds based on the predictability and maturity of their build environment.

\subsection{Relevance Varies by Failure Type}
Manual inspection revealed that noise levels differ across failure categories. Compilation and dependency errors tend to produce dense diagnostic traces, whereas permission or configuration issues often yield boilerplate output. 
This heterogeneity suggests that a single global reduction ratio may not be optimal for all workflows. Adaptive reduction strategies that condition on failure type, pipeline stage, or project domain could further improve both precision and recall of relevant lines. For example, during dependency resolution, a higher retention threshold may capture transient network or package errors, whereas build or test phases can tolerate stronger filtering once patterns of compiler diagnostics are learned.  Incorporating such context-aware adaptation would allow \textit{LogSieve} to evolve from a static pre-filter into a dynamic component that continuously learns relevance distributions as CI pipelines mature.

\subsection{Empirical Comparison with LogZip}
RQ4 demonstrates that \textit{LogSieve} preserves semantic fidelity far better than \textit{LogZip} at equivalent reduction levels. For explanation (Prompt~1), \textit{LogSieve} improves similarity to full-log responses by roughly 20--35 percentage points across CosSim, ROUGE-1/ROUGE-L, and BLEU (0.93 vs.\ 0.70; 0.73 vs.\ 0.51; 0.47 vs.\ 0.26; 0.42 vs.\ 0.09), by 11 points on BERTScore (0.79 vs.\ 0.68), and by 52 points on GPTScore (0.93 vs.\ 0.41). For categorization (Prompt~2), exact-match accuracy increases from 0.20 to 0.80; the \emph{Random} baseline reaches 0.70, reflecting higher surface similarity but lower categorical fidelity. These gaps indicate that structure-first compression can discard contextual cues essential for LLM reasoning, whereas RCA-aware filtering maintains interpretability while reducing token cost.

\subsection{Position Relative to Existing Techniques}
Traditional techniques such as \textit{LogZip}~\cite{zhu2019logzip} and
\textit{LogShrink}~\cite{li2024logshrink} focus on compression or deduplication for
storage efficiency but overlook semantic relevance. \textit{Drain}~\cite{he2017drain} and
\textit{Spell}~\cite{du2017spell} abstract syntax without modeling task importance, while
\textit{LoFI}~\cite{huang2024lofi} integrates LLMs into reduction but incurs inference overhead.
In contrast, \textit{LogSieve} performs semantics-aware filtering \emph{before} inference,
complementing structural tools in token-sensitive workflows.

\begin{table}[t]
\centering
\scriptsize
\setlength{\tabcolsep}{1pt}
\caption{Conceptual comparison of \textit{LogSieve} with log compression, structural parsing, and LLM-in-the-loop techniques.}
\begin{tabular*}{\columnwidth}{@{\extracolsep{\fill}} lccccc}
\toprule
\textbf{Property} & \textbf{LogSieve} & \textbf{LogZip} & \textbf{Drain} & \textbf{Spell} & \textbf{LoFI} \\
\midrule
Semantic / task-aware relevance  & \cmark & \xmark & \xmark & \xmark & \cmark \\
Structural parsing or templates  & --     & \cmark & \cmark & \cmark & --     \\
Token reduction before LLM       & \cmark & \xmark & \xmark & \xmark & \xmark \\
Needs LLM during reduction       & \xmark & \xmark & \xmark & \xmark & \cmark \\
Online / streaming capable       & --     & --     & \cmark & \cmark & --     \\
Main objective                   & Semantics & Compression & Parsing & Stream parsing & Fault signals \\
Evaluated on CI logs             & \cmark & \xmark & \xmark & \xmark & \xmark \\
LLM inference overhead           & Low    & Low    & Low    & Low    & High  \\
\bottomrule
\end{tabular*}
\label{tab:technique-comparison}
\end{table}

\subsection{Sustainability and Cost Implications}
RCA-aware pre-reduction directly lowers resource use for LLM inference and CI infrastructure. We do not directly measure energy consumption or emissions; instead, we report token reductions and discuss proportional cost/energy implications using established Green AI models. In our dataset, the average prompt size fell from 26{,}543 to 13{,}355 tokens per run ($-40\%$), and the number of lines dropped from 732 to 381 ($-42\%$). Because inference cost and latency scale with input tokens, these reductions  translate to immediate savings in computation and can yield proportional reductions in energy use (and associated emissions) ~\cite{Koenig2025,Naumann2011,schwartz2020green,henderson2020towards,wu2022sustainable}.

\begin{table}[t]
\centering
\setlength{\tabcolsep}{2pt}
\caption{Resource summary (means across 20 repositories). $\Delta$ reports the macro-averaged percentage reduction; absolute means are shown in the first two columns. Token reduction dominates inference-cost and latency improvements.}
\begin{tabular*}{\columnwidth}{@{\extracolsep{\fill}} lccc}
\toprule
 & \textbf{Full} & \textbf{LogSieve} & \textbf{$\Delta$~(\%)} \\
\midrule
Mean input tokens / run & 26{,}543 & 13{,}355 & $-40$ \\
Mean lines / run        & 732      & 381       & $-42$ \\
\bottomrule
\end{tabular*}
\label{tab:sustainability-summary}
\end{table}

\paragraph{Parametric cost model.}
Following prior work on Green AI and energy-aware computation~\cite{schwartz2020green,henderson2020towards,wu2022sustainable}, we model inference cost as a linear function of token usage:
\[
\mathrm{Cost}=\frac{T_{\mathrm{in}}}{1000}\,p_{\mathrm{in}}+\frac{T_{\mathrm{out}}}{1000}\,p_{\mathrm{out}}.
\]
Pre-reduction primarily decreases $T_{\mathrm{in}}$. With $T_{\mathrm{full}}{=}$26{,}543, $T_{\mathrm{red}}{=}$13{,}355, and $T_{\mathrm{rem}}{=}$13{,}188,
\[
\Delta \mathrm{Cost}_{\mathrm{in}}=\frac{T_{\mathrm{rem}}}{1000}\,p_{\mathrm{in}}\approx 13.188\,p_{\mathrm{in}} \;\text{per run.}
\]
If response length remains similar (as in categorical tasks), $\Delta T_{\mathrm{out}}$ is negligible; otherwise the full expression applies.

\paragraph{Energy and carbon perspective.}
Following established energy–carbon estimation models for machine learning~\cite{strubell2020energy,patterson2021carbon,patterson2022carbon}, we estimate emissions as
\[
\Delta \mathrm{CO}_2=\Delta E\cdot CI_{\text{grid}},\qquad \Delta E\propto T_{\mathrm{in}}.
\]

A 40~\% token reduction therefore yields a proportional decrease in energy use for LLM-assisted CI analysis while also reducing bandwidth and storage requirements for log artifacts. These efficiency gains require no model retraining and complement structural tools by filtering semantically irrelevant content before inference.

Beyond immediate cost and energy considerations, such efficiency gains are particularly relevant in long-lived software ecosystems. Prior empirical studies have shown that ecosystems can undergo periods of slowdown and abandonment, during which maintenance and diagnostic responsibilities increasingly fall on a shrinking set of contributors~\cite{amit2025understanding}. In these contexts, reducing routine CI analysis overhead becomes critical, as developers must diagnose failures and maintain pipelines under constrained time and resource budgets. Lightweight, semantics-aware pre-reduction helps lower this burden without requiring changes to downstream tools or CI workflows.

\subsection{Integrative Reflection}
Across all analyses, \textit{LogSieve} demonstrates that RCA-aware reduction can bridge the gap between traditional log compression and LLM-based understanding. By retaining semantics rather than syntax, it enables efficient, scalable, and environmentally conscious automation within CI workflows.

\section{Threats to Validity}
\label{sec:Threats_to_validity}

As with most empirical studies, our findings are subject to several threats to validity. We outline potential limitations and describe the measures taken to mitigate them.

\paragraph{Internal Validity}
Potential bias arises from the manual labeling of log lines. Labels were assigned using heuristic criteria (e.g., prioritizing error messages and de-emphasizing setup steps), reflecting how reduction rules might be constructed in practice. To mitigate subjectivity, two co-authors labeled independently, disagreements were resolved with a third, and analyses emphasized aggregate trends. A residual concern is \emph{heuristic anchoring}—the possibility that labelers and models rely on similar cues. We mitigate this through publicly stated heuristics, transparent inter-rater reliability reporting (Cohen’s~$\kappa{=}0.80$ in RQ1), and evaluation on held-out folds. 

Cross-validation also introduces potential for \emph{data leakage} if lines from the same workflow appear across folds, which could inflate accuracy. We used stratified 10-fold cross-validation to maintain class balance, but future work will employ group-stratified CV by repository to further reduce this risk. Hyperparameter tuning on test folds can overstate performance; nested CV or a held-out validation set would be stricter. Finally, embeddings were treated as fixed features—fine-tuning or domain-adaptive pretraining could change results and should be explored in replication.

\paragraph{External Validity}
Our dataset covers 20 open-source Android projects using GitHub Actions (failed runs). Findings may not generalize to other CI ecosystems (e.g., GitLab CI, Azure DevOps), programming languages, project scales, or log types (system, server, telemetry), nor to proprietary pipelines. The focus on failed runs and GitHub’s 90-day retention window may bias toward recent, failure-heavy activity. Similarly, reliance on a single LLM (GPT-4o) limits generalizability; replication with other models and human judgments would strengthen conclusions. Broader validation across CI platforms, languages, and failure modes will enhance generalizability.

\paragraph{Construct Validity}
We operationalize relevance using a binary retain/remove label, which simplifies how lines contribute to different diagnostic or summarization tasks. Prompts modeled two specific tasks (failure explanation and categorization); other CI analyses (e.g., repair suggestion) may require different evidence. Similarity metrics (Cosine, BERTScore, ROUGE, BLEU, GPTScore) approximate semantic fidelity but are imperfect proxies for developer utility—high textual similarity does not necessarily translate to faster debugging. We do not directly measure operational energy consumption or emissions; sustainability implications are inferred from token reductions under proportional energy–token assumptions used in prior Green AI models. GPTScore, as an LLM-based judge, may also introduce model-specific bias; triangulating with lexical metrics and exact-match accuracy partly mitigates this risk. Baseline configuration choices (e.g., \textit{LogZip} parameters, random seeds) may also influence outcomes. We used default settings to reflect typical usage and report distributional summaries (median, min, max) to reduce interpretation bias.

\paragraph{Conclusion Validity}
Our evaluation included 20 repositories, sufficient for an initial feasibility study but not for broad statistical generalization. While effect sizes are large, variance remains. Future iterations will include confidence intervals, bootstraps, and per-repository analyses. We also plan paired significance tests (e.g., McNemar) for classification agreement and random-seed robustness checks to validate reproducibility.

\paragraph{Reproducibility and Evolution Threats}
LLM services evolve (e.g., model snapshots, safety filters, tokenizers), which may alter outputs over time. We fixed prompts, temperature, and tokenizer (\texttt{tiktoken}) and will release all artifacts—prompts, code, and labels—for replication. We recommend pinning model versions, logging random seeds, and exporting raw responses to mitigate drift. For automation experiments, training code and cross-validation splits will be released to support long-term reproducibility.

\paragraph{Ethical and Privacy Considerations}
We used public open-source CI logs, which may still contain incidental identifiers. We avoided storing PII and will provide redaction guidelines in our replication package. For private CI environments, organizations should apply privacy filters and enforce access control. By releasing our code, labels, and prompts, we aim to support open, transparent, and ethical research in automated CI analytics.

\section{Conclusion}
\label{sec:Conclusion}
We studied CI logs from 20 open-source Android projects on GitHub Actions (failed runs) and introduced \textit{LogSieve}, a RCA-aware, semantics-preserving log reduction technique that filters low-information lines while retaining content relevant to downstream LLM-based analysis. Across repositories, \textit{LogSieve} removed on average 42\% of lines and 40\% of tokens (RQ1). Despite these reductions, semantic fidelity remained
high for explanation responses (Cosine = 0.93, GPTScore = 0.93), and failure categorization matched full-log labels in 80\% of cases (RQ2). Embedding-based classifiers automated relevance labeling with near-human accuracy (up to 97\%; RQ3). Compared with structure-first compression approaches such as \textit{LogZip}, \textit{LogSieve} preserved substantially more RCA-relevant semantics for both explanation and categorization (e.g., Cosine 0.93 vs.\ 0.70; exact-match 80\% vs.\ 20\%; RQ4). By reducing token budgets before inference, \textit{LogSieve} advances sustainable, interpretable CI automation without modifying model weights or prompts.

\paragraph{Future Work.}
Future research will extend \textit{LogSieve} in several directions. We plan to benchmark it against additional reduction and parsing techniques (e.g., \textit{Drain}~\cite{he2017drain}, \textit{Spell}~\cite{du2017spell}, \textit{LoFI}~\cite{huang2024lofi}) on shared datasets and standardized downstream tasks to better contextualize trade-offs between structure-first and task-aware approaches. We also aim to automate relevance detection using heuristics and weak supervision, support adaptive reduction based on token budgets and failure types, and generalize to other log domains such as server or telemetry data. Beyond method extensions, we will evaluate \textit{LogSieve} with additional LLMs to assess model-agnostic robustness and quantify end-to-end energy and carbon savings in real deployments. We plan to explore hybrid pipelines that integrate \textit{LogSieve} with structural parsers (e.g., \textit{Drain}) to balance interpretability, scalability, and sustainability. Finally, we intend to complement LLM-based evaluation with a practitioner survey to assess whether the reduced logs are also more  meaningful and interpretable for humans. This will help establish whether the task-aware reduction principles demonstrated here generalize beyond root-cause analysis to other CI and software-maintenance tasks, uniting human and model-centric perspectives on sustainable automation.

\paragraph{Artifacts and Replicability.}
We provide a publicly available replication package supporting this study. The artifacts are hosted on Figshare and can be accessed via \url{https://doi.org/10.6084/m9.figshare.30811382}. The package enables replication of the analyses reported in this paper and supports further investigation of task-aware CI log reduction, including RCA-focused analyses.

\begin{acks}
Funding: We acknowledge the support of the Natural Sciences and Engineering Research Council of Canada (NSERC): [RGPIN-2021-03969] and [RGPIN-2025-05897].
\end{acks}

\balance
\bibliographystyle{ACM-Reference-Format}
\bibliography{References}

\end{document}